\documentstyle[12pt]{article}
\begin{document}
\newtheorem{thm}{Theorem}
\newtheorem{cor}{Corollary}
\newtheorem{Def}{Definition}
\newtheorem{lem}{Lemma}
\begin{center}
{\large \bf Equations of motion in General Relativity and Quantum
Mechanics } \vspace{5mm}

Paul O'Hara
\\
\vspace{5mm}
{\small\it
Dept. of Mathematics\\
Northeastern Illinois University\\
5500 North St. Louis Avenue\\
Chicago, Illinois 60625-4699.\\
\vspace{5mm}
email: pohara@neiu.edu \\}
\end{center}
\vspace{10mm}
\begin{abstract}
In a previous article a relationship was established between the
linearized metrics of General Relativity associated with geodesics
and the Dirac Equation of quantum mechanics. In this paper the
extension of that result to arbitrary curves is investigated. A generalized
Dirac equation is derived and shown to be related to the Lie
derivative of the momentum along the curve. In addition,the
equations of motion are derived from the Hamilton-Jacobi equation
associated with the metric and the wave equation associated with the
Hamiltonian is then shown not to commute with the Dirac operator.
Finally, the Maxwell-Boltzmann distribution is shown to be a
consequence of geodesic motion.
\newline

\end{abstract}

\section {Introduction}

In a recent paper Frank Tipler \cite{tip} gives a derivation of the
Schrodinger equation using the Hamilton-Jacobi principle of action.
In doing so, he is able to transpose Erwin Schrodinger's ``{\it
purely} formal procedure''\cite{schr} of replacing $\frac{\partial
W}{\partial t}$ in the Hamilton-Jacobi equation with
$\pm\frac{h}{2\pi i}\frac{\partial }{\partial t}$ with a more direct
derivation of the wave equation. Likewise, in a paper by
Marie-Noelle Celerier and Laurent Nottale \cite{marie} in 2003, the
Dirac equation is derived by taking the square root of the Klein-Gordan
equation using the ``bi-quaternionic action'' associated with
geodesic motion and the Hamilton-Jacobi principle of action. Other
approaches not directly based on the use of the action principle
have also been used. For example, Ng and Dam used a geometrical
derivation of the Dirac equation by exploiting ``rotational
invariance'' and ``the explicit use of the spin-$\frac12$ property
of $\psi$''\cite{ng}, while Martin Rivas quantized a Poincare
invariant Hamiltonian in which the spin angular momentum of the
particle is constant with respect to the center of mass observer to
derive the Dirac equation.\cite{rivas}\vskip 5pt This paper contains
parallels with the approach of Tipler, and Celerier and Nottale in
that it makes use of the Hamilton Jacobi equation but it also
differs in that it emphasizes the fundamental role of the metric in
enabling us to derive two equations associated with non-geodesic
motion: one a generalized Dirac equation, which captures the
kinematics, and the other an equation of motion describing the
dynamics.

Throughout the paper, $({\cal M},g)$ will denote a connected four
dimensional Hausdorff manifold, with metric $g$ of signature -2. At
every point $p$ on the space-time manifold $\cal M$ we erect a local
tetrad $e_0(p), e_1(p), e_2(p), e_3(p)$ such that a point $x$ has
coordinates $x=(x^0,\ x^1,\ x^2,\ x^3)=x^ae_a$ in this tetrad
coordinate system, while the spinor $\Psi$ can be written as $\Psi
=\psi^ie_i(p)$, where $\psi^i$ represent the coordinates of the
spinor with respect to the tetrad at $p$. This is permissable since
a spinor is also an element of a vector space.  In particular, when
each $\psi^i$ is equal for each $i$ then we can write $\Psi=\psi
\xi$\, where the spinor $\xi=(e_i)$ and $||\Psi||=1$. When this occurs, we say
that $\psi$ is a scalar field. Such a scalar field will occur when
$\Psi$ is parallel transported along a geodesic. For example, in the
case of the Dirac equation, the plane wave solution can be written
as
\begin{eqnarray} \Psi(x)=e^{-ipx/\hbar}u_+(p), \end{eqnarray}
where $\psi = e^{-ipx/\hbar}$ and $u_+(p)=\left(\begin{array}{c}
                                            u_1 \\
                                            u_2
                                          \end{array}\right)
$ and allows for two independent solutions for each momentum $p$. Also
in this case, the spinor $u(p)$ remains constant along a
geodesic. This reflects Rivas result in that the derivation of
the Dirac equation presupposes the motion of a particle ``in a plane
orthogonal to $S$, which is constant in this frame,'' \cite{rivas}
where $S$ refers to spin-angular momentum. In terms of quantum
mechanics each constant can be identified with a quantum number.
Also at each point $p$ we can establish a tangent vector space
$T_p(\cal M)$, with basis $\{\partial_0,
\partial_1,
\partial_2, \partial_3\}$ and a dual 1-form space, denoted by $T^*_p$ with
basis $\{dx_0,dx_1,dx_2,dx_3\}$ at $p$, defined by
\begin{eqnarray}
dx^{\mu}\partial_{\nu}\equiv
\partial_{\nu} x^{\mu}=\delta^{\mu}_{\nu}.
\end{eqnarray}
We refer to the basis $\{dx^0, dx^1, dx^2, dx^3\}$ as the basis of
one forms dual to the basis $\{\partial_0, \partial_1, \partial_2,
\partial_3\}$ of vectors at $p$.

With notation clarified, we note that in a previous article
\cite{ohara}, the Dirac equation was derived as a dual of a
linearized metric tensor for geodesics. The linkage was accomplished in a
natural way by associating a generalized Dirac equation with those
operators which are duals of differential one-forms, obtained by
linearizing the metric tensors of General Relativity (expressed locally as
a Minkowski metric). Specifically, in that paper, if
\begin{eqnarray} ds^2=g_{\mu \nu}dx^{\mu}dx^{\nu}=\eta_{ab}dx^adx^b
\end{eqnarray}
where $a$ and $b$ refer to local tetrad coordinates and $\eta$ to a
rigid Minkowski metric tensor of signature -2, it was shown that a matrix
$\tilde{ds}\equiv\gamma_{a}dx^{a}$ with eigenvalue $ds$ could be associated with the metric tensor
$\tilde{ds}\otimes \tilde{ds}$ (which from now on we will denote by $ds^2$) such that $\{\gamma_a,\gamma_b\}=2\eta_{ab}$.

In addition, by noting that the $\tilde{ds}$ matrix is the dual
of the expression $\tilde{\partial}_s\equiv \gamma^a\frac{\partial}
{\partial x^a}$ this enabled us to associate a generalized
Dirac equation
\begin{equation} \tilde{\partial}_s \psi\xi\equiv
\gamma^a\frac{\partial \psi} {\partial x^a}\xi=\frac{\partial
\psi}{\partial s}\xi,
\end{equation}
with the motion of a particle along a geodesic, with a
fixed spinor $\xi$.

This paper is an extension of that previous work. First,
we re-investigate this result by showing that the
existence of a Hamilton-Jacobi function associated with a family of space
filling curves is equivalent to defining the Dirac equation at almost every point on the manifold. Indeed, this association
also suggests a deeper understanding of the term ``wave-particle duality.''
Secondly, we show that the existence of such functions also determines the existence of both quantum and classical ideal gases defined in terms of
the Bose-Einstein and Boltzmann statistics respectively. Thirdly,
we introduce accelerations and then show that the integrability of
the Hamilton-Jacobi function determines a coherent set of natural motions and also implies the non-commutativity of the Hamiltonian and action operators  associated with non-geodesic motion.

\section{Generalized Dirac Equation}

The generalized Dirac equation defined above relies on the
definition of the four-momentum in special relativity and upon the
fact that $\tilde{\partial}_s\psi$ and $\tilde{ds}$ are parallel
along geodesics, and consequently by the chain rule their product is
$\frac{d \psi}{ds}ds$. In contrast, when accelerations are
introduced we will find that in general
\begin{eqnarray}\frac{\tilde{ds}}{ds}.\tilde{\partial}_s\psi&=&\frac12\left\{\frac{\tilde{ds}}{ds},\tilde{\partial}_s\psi\right\}+\frac12
\left[\frac{\tilde{ds}}{ds},\tilde{\partial}_s\psi\right]\\
&=&\frac{d \psi}{d s}+\frac{\vec{ds}}{ds}\wedge\vec{\frac{\partial
\Psi}{\partial s}}.
\end{eqnarray}
In this regard, note that equation (5) represents an operator equation and the four $\gamma$ matrices used to define ${\tilde ds}=\gamma^adx_a$
and ${\tilde \partial_s \psi}=\gamma^a\frac{\partial \psi}{\partial x_a}$ are a spinor representation of four unit vectors constituting a local Minkowski tetrad at any point on the curve.
Equation (6), on the other hand, is written in the usual vector notation. The two equations can be identified by noting that the anti-commutator
and commutator relationships in (5) associated with the spinor tetrad, define a dot product and a cross product respectively,
which can can also be expressed in standard vector notation as in equation (6).
Direct calculation of the dot product of $\tilde{ds}$ and its dual
$\tilde{\partial_s}\psi$ or equivalently of ${\vec{dx^a}}$ and $\vec\frac{\partial \psi}{dx_a}$ gives $\frac{d \psi}{ds}ds$.

This latter term can also be directly related to the Hamilton-Jacobi
characteristic function, which in turn can be associated with a coherent set of natural
motions \cite{Synge}. Indeed, to remove any ambiguity, we begin with
the following definitions:
\begin{Def} A function $W=\int_{\sigma(\lambda)}\left ({\mathbf{p}}{\frac{\mathbf{dx}}{d\lambda}}-H\frac{dt}{d\lambda}\right )d\lambda$
is called a Hamilton-Jacobi function if the integral is path independent for all curves in $\{\sigma(\lambda) \in ({\cal M},g)\}$ and
$\frac{\partial W}{\partial t}=-H(x_1,x_2,x_3,t, \frac{\partial W}{\partial x_1}, \frac{\partial W}{\partial x_2},\frac{\partial W}{\partial x_3})$ defined with respect to a local tetrad .
\end{Def}
Equivalently, we can say that $dW={\mathbf{pdx}}-Hdt$ is an exact differential.\\

We begin by focusing on the kinematics associated with the function $W$ and by showing how it can be related to the Dirac equation. This leads to the following lemmas and corollaries:

%
%

%

\begin{lem}Let $\psi(W(x^a))$ be a differentiable function of $W$ and let $\psi^{\prime}=\frac{d\psi}{dW}$. $W=-\int \eta^{ab} p_adx_b=\int \textbf{pdx}-Hdt$ is the Hamilton-Jacobi function, iff $\psi(W)=\int \textbf{p}^*\textbf{dx}-H^*dt$ is also a Hamilton-Jacobi function.
\end{lem}
{\bf Proof:}$(\Rightarrow)$ First note $p^*_a=\psi^{\prime}p_a$, from the differentiability of $\psi(W)$, and $p_o=H$.
Since $W$ is a Hamilton-Jacobi function:
\begin{eqnarray*} \frac{{\partial \psi(W)}}{{\partial t}}&=&\psi^{\prime}\frac{{\partial W}}{{\partial t}}\\
&=&-\psi^{\prime}H(x_1,x_2,x_3, t,p_1,p_2,p_3)\\
&=&-\psi^{\prime}H(x_1,x_2,x_3, t,p^*_1/\psi^{\prime},p^*_2/\psi^{\prime},p^*_3/\psi^{\prime})\\
&=&-H^*(x_1,x_2,x_3, t, p^*_1, p^*_2, p^*_3).
\end{eqnarray*}
Therefore, $\psi(W)$ is a Hamilton-Jacobi function.
\\

\noindent $(\Leftarrow)$ Conversely if $\psi(W)$ is a Hamilton-Jacobi function then
\begin{eqnarray*} \frac{{\partial W}}{{\partial t}}&=&\frac{{\partial \psi(W)}}{{\partial t}}/\psi^{\prime}\\
&=&-H^*(x_1,x_2,x_3, t,p^*_1,p^*_2,p^*_3)/\psi^{\prime}\\
&=&-H(x_1,x_2,x_3, t,p^*_1/\psi^{\prime},p^*_2/\psi^{\prime},p^*_3/\psi^{\prime})\\
&=&-H(x_1,x_2,x_3, t, p_1, p_2, p_3).
\end{eqnarray*}
Therefore, $W$ is a Hamilton-Jacobi function.
\\

\noindent In addition, from equations (5) and (6) we obtain the following Lemma which gives a necessary and sufficient condition for the existence of
such Hamilton-Jacobi functions.
\begin{lem}
Let $\psi(W({\mathbf{x}},t))$ be a differentiable function and $\{\sigma(\lambda)\}$ a family of curves on the manifold with unit tangent vectors $\frac{\tilde{ds}}{ds}$ with respect to a local tetrad then $\tilde{ds}.\tilde{\partial}_s \psi(W)$ is an exact differential iff  $\psi(W)$ is a Hamilton-Jacobi function such that $p^*_a=\frac{d\psi}{ds}\frac{dx_a}{dt}=\frac{\partial \psi_{\|}(W)}{\partial x^a}$, where $\psi(W)=\psi_{\|}(W) + \psi_{\bot}(W)$.
\end{lem}
{\bf Proof:} From equations (5), if $\frac{\tilde{ds}}{ds}.\tilde{\partial}_s \psi(W)$
is an exact differential then $ [\frac{\tilde {ds}}{ds}, \tilde \frac{\partial \psi(W)}{\partial s}]$ is also an exact differential. This means that
$\psi(W)=\psi_{\|}(W)+\psi_{\bot}(W)$ where $[\frac{\tilde {ds}}{ds}, \tilde \frac{\partial \psi_{\|}(W)}{\partial s}]=0$ and
$\psi_{\bot}(W)=c_ot+c_1x_1+c_2x_2+c_3x_3$,  with $c_o,c_1,c_2,c_3$ being constants.\\

\noindent Also, $[\frac{\tilde {ds}}{ds}, \tilde \frac{\partial \psi_{\|}(W)}{\partial s}]=0$ means $\frac{{dx^a}}{d\lambda}$is parallel (for any parameter $\lambda$ including the curve length $s$) to $\vec \frac{\partial \psi_{\|}(W)}{\partial s}$, and $\exists g(\lambda)$ such that $p^{*}_{\| a}\equiv g(\lambda)\frac{dx_a}{d\lambda}=\frac{\partial \psi_{\|}(W)}{\partial x^a}$.\newline

\noindent Also given $\frac{\tilde{ds}}{ds}.\tilde{\partial}_s \psi_{\|}(W)$ is an exact differential and denoting $x_0=t$, gives
\begin{eqnarray*}\frac{\tilde{ds}}{ds}.\tilde{\partial}_s\psi_{\|} (W)&=&\frac{\partial \psi_{\|}(W)}{\partial x^1}\frac{dx^1}{ds}+
\frac{\partial \psi_{\|}(W)}{\partial x^2}\frac{dx^2}{ds}+
\frac{\partial \psi_{\|}(W)}{\partial x^3}\frac{dx^3}{ds}+\frac{\partial \psi_{\|}(W)}{\partial t}\frac{dt}{ds}\\
&=&\frac{d\psi_{\|}(W)}{ds}.
\end{eqnarray*}
On substituting $g(s)\frac{dx_a}{ds}=\frac{\partial \psi_{\|}(W)}{\partial x^a}$ and noting that $\frac{dx^a}{ds}\frac{dx_a}{ds}=1$ gives\newline $g(s)=\frac{d\psi_{\|}(W)}{ds}$. It follows that
\begin{equation}\left(\frac{\partial \psi_{\|}(W)}{\partial t}\right)^2=(p^{*1}_{\|})^2+(p^{*2}_{\|})^2+(p^{*3}_{\|})^2+ \left(\frac{d\psi_{\|}(W)}{ds}\right)^2=\left(H^*_o\left(x^a, p^{*1}_{\|}, p^{*2}_{\|}, p^{*3}_{\|}\right)\right)^2.\end{equation} Therefore $\psi_{\|}(W)$ is a Hamilton-Jacobi function, as also is $\psi(W)=\psi_{\|}(W)+\psi_{\bot}(W)$ with $p^{*}_a=p^{*}_{\| a}+c_{a}$.
\\

\noindent Conversely, given $p^*_a=\frac{d\psi}{ds}\frac{dx_a}{dt}+c_a=\frac{\partial \psi_{\|}(W)}{\partial x^a}+\frac{\partial \psi_{\bot}(W)}
{\partial x^a}=\frac{\partial \psi(W)}{\partial x^a}$ then $\left[\tilde{ds},\tilde{\partial}_s\psi_{\|}\right]=0$
and since $\psi(W)$ is a Hamilton-Jacobi function, it follows from Equation (5) and the definition of $\psi_{\bot}$ that
$$\tilde{ds}.\tilde{\partial}_s \psi_{\|}(W)=d \psi_{\|}(W)\qquad \textrm{and}\qquad \left[\frac{\tilde {ds}}{ds}, \tilde \frac{\partial \psi_{\bot}(W)}{\partial s}\right]$$
are both integrable.  Therefore,$\tilde{ds}.\tilde{\partial}_s \psi(W)$ is an exact differential.
The result follows.

\begin{cor} If $\psi(W)=W=-m_os+k$ is a Hamilton-Jacobi function for each fixed $k$ then $g(s)\equiv m_o$ is called the rest mass associated with linear motion in the plane $W(s,k)$, and
$$H^2=p^2_1+p^2_2+p^2_3+m_o^2.$$
\end{cor}
{\bf Proof:} Follows from Equation (7).
\\

We now show that for the family of curves defined in Lemma 2, we can always construct a Hamilton-Jacobi function locally at each point on the curve.
\begin{cor}
Let $ds^2=dx^adx_a$ define a metric tensor locally along a piece-wise smooth congruence of curves,$\{\sigma(\lambda)=x^a(\lambda)\}$, which fill coordinate space with only one curve passing through each point, then a Hamilton-Jacobi function can be constructed such that
\begin{equation} W=\int_{\sigma} m\frac{ds}{d\lambda}ds =\int
\textbf{pdx}-m\frac{dt}{d\lambda}dt,\qquad \textrm{ where}\quad
p^a=m\frac{dx^a}{d\lambda},\ m=m_o\frac{d\lambda}{ds}.\end{equation}
Note $\frac{\partial W}{ds}=m_o$ is called the rest mass.
\end{cor}
{\bf Proof:}
At every space-time point in a gravitational field, by the Principle of Equivalence, an inertial frame can be constructed. Now, define $p^a=m\frac{dx^a}{d\lambda},\ m=m(s)\frac{d\lambda}{ds}$,
where $m(s)=m_o$ is a constant. Then $dW\equiv m\left(\frac{ds}{d\lambda}\right)^2d\lambda=m_ods$ is clearly an exact differential,
such that $$W=m_os+k=m_o\sqrt{x^2_1+x^2_2+x^2_3-t^2}+k, \qquad k \ \textrm{constant}$$
in a neighborhood of any point. Therefore $\tilde{\partial_s} W=\gamma^a\frac{\partial W}{\partial x^a}$ exists almost everywhere
and  $dW=\tilde{ds}.\tilde{\partial}_s \psi(W)$. It follows from Lemma 2 that $W$ is a Hamilton-Jacobi function.
\\

\noindent Remark: The corollary has shown the existence of $W$ at every
point on the manifold by showing that $W$ can be constructed in
terms of a local tetrad as a linear function of the arc length, s.
In effect, this means that we are defining the Hamilton-Jacobi function
along a geodesic passing through the point $x_o$. In practice, it is sufficient to integrate $\frac{\partial W}{\partial
x^a}=m_o\frac{dx^a}{ds}$ to obtain $W$, as in the example below. Moreover, although $p_o=p(x_o)$
is a constant along the geodesic passing through $x_o$, this does
not mean that $p^a= \frac{\partial W}{\partial x^a}$ is constant
along an arbitrary curve. Indeed, this latter condition only follows, if
motion is along a geodesic.
\newline

\noindent \textbf{Example:} As an example of the corollary, consider a motion of a particle of rest
mass $m_o$ fired into the air without resistance. The Equation of
Motion for such a projectile in Minkowski space are given by
\begin{equation} {\mathbf{F}}=-m_og\mathbf{j},\end{equation}
or equivalently, denoting $\frac{dx}{ds}=\dot{x}$,
\begin{eqnarray} m_o\ddot{x}=0\qquad m_o\ddot{y}=-m_og. \end{eqnarray}
Solving the equations gives \begin{equation} x=x_0+u_xs \quad
\textrm{and} \qquad y=y_0+u_ys-\frac{1}{2}gs^2.\end{equation}

However, in terms of the corollary above, we begin not with equations of motion (9)
but with the line increments associated with the kinematics described by equation (11). Specifically,
\begin{eqnarray*}
-ds^2 &=& dx^2+dy^2-dt^2\\
\textrm{iff}\qquad -m_o ds &=& m_o\dot{x}dx+m_o\dot{y}dy-m_o\dot{t}dt\\
 \textrm{iff}\qquad -m_o ds &=& m_o
u_xdx+m_o(u_y-gs)dy-m_o\dot{t}dt.\end{eqnarray*} If we now let
$dW=-m_ods$ along a geodesic and require that $W$ be an integrable function passing through $(x_0,y_0,t_0)$,
we obtain $W=m_ou_xx+m_o(u_y-gs)y-m_oc^2\dot{t}t+w_o$. Indeed, on taking
partial derivatives, we find
$$\frac{\partial W}{\partial x}=m_ou_x=p_x,\qquad \frac{\partial W}{\partial y}=m_o(u_y-gs)=p_y\qquad
\left(\frac{\partial W}{\partial t}\right)=-m_o\dot{t}$$ and
$$ \left(\frac{\partial W}{\partial t}\right)^2 =m_o^2+p^2_1+p^2_2+p^2_3=H^2.$$
Therefore, H is a Hamilton-Jacobi function.
\newline

In effect, Lemma 2 establishes a relationship between Hamilton-Jacobi functions and the commutator relationship, $[\tilde {ds},\tilde {\partial_s}\psi(W)]$
applied to Equation (5) (or equivalently Equation (6)). We now use the same commutator relationship to establish another important property relating
the dual operator $\tilde{\partial}_s$ and the metric operator $\tilde{ds}$ associated with the increments along a curve. Indeed, Equation (5) could be
described as the most general form of a ``wave-equation'' associated with a curve in space-time. However, we put the expression ``wave-equation'' in quotes
to emphasize that $\psi(W)$ is not necessarily a wave-function of quantum mechanics. For the moment, $\psi$ they can be any $C_1$ function defined on the
manifold. In a previous paper, we have noted that $\psi$ can only be interpreted as a quantum wave function when further restrictions are imposed
on the function space, such as requiring that it be an $L^2$ function. Intuitively, we could think of $\tilde{\partial}\psi(W(s))$ as a wave associated with the vibration of a curve $\sigma(s)$ in space-time, whose tangent is $\tilde{ds}$, with respect to a local tetrad coordinate system.
This leads to the following lemma:
\begin{lem} If $\psi(W)$ is a Hamilton-Jacobi function such that $[\tilde{\partial_s}W,\tilde{ds}]=0$ then there exists a simultaneous eigenfunction $\xi$ such that
\begin{equation}(\tilde{\partial_s}\psi)\xi(p)=\partial\psi_s\xi(p),\end{equation}where $\partial\psi_s=\frac{\partial \psi}{\partial s}$, which in the
case of geodesic motion reduces to
\begin{equation}\tilde{\partial_s} \Psi
=\frac{d\Psi}{ds},\qquad \textrm{where} \qquad \Psi=\psi \xi.
\end{equation}
\noindent \textbf{Remark:} $(\tilde{\partial_s}\psi)\xi=
\tilde{\partial_s}\Psi$ in general, since $x$ is independent of $p$
in phase space. However, $\frac{d\Psi}{ds}\neq \psi^{\prime}(p)\xi$
unless motion is along a geodesic.
\end{lem}
{\bf Proof:}
First note that $[\tilde{\partial_s}W,\tilde{ds}]=0$ implies $[\tilde{\partial_s}\psi,\tilde{ds}]=[\psi^{\prime}(W)\tilde{\partial_s}W,\tilde{ds}]=0$. Therefore, there exists simultaneous
eigenvectors $\xi=\xi(p)$ such that $\tilde{ds}\xi=ds\xi$ and
$(\tilde{\partial_s}\psi)\xi=\gamma^a p^*_a\xi(p)=\gamma^ap_a \psi^{\prime}
\xi(p)=mc\psi^{\prime}(p)\xi(p)=(\partial_s\psi)\xi(p)$.\newline Also,
$\xi(p)$ is constant along a geodesic and therefore
$$\tilde{\partial_s} \Psi =\frac{d\Psi}{ds},\qquad \textrm{where}
\qquad \Psi=\psi \xi.$$ The result follows.
\\
\begin{cor}Let $ds^2=dx^adx_a$ define a metric tensor on a manifold and $p^a=m(s)\frac{dx^a}{ds}$ along a curve.
Then there exists a Hamilton-Jacobi function $\psi(W)$, and a vector $\xi(p)$ such that
$$(\tilde{\partial_s}\psi)\xi(p)=(\partial_s\psi)\xi(p).$$
\end{cor}
{\bf Proof:} By Cor.2 it is sufficient to define $p_a=\frac{\partial W}{\partial x^a}$, and solve for $W$ to obtain a Hamilton-Jacobi equation. It now follows that
\begin{equation}[\tilde{\partial_s}W,\tilde{ds}]=[\gamma^a\frac{\partial
W}{\partial x^a},
\gamma^adx_a]=[\gamma^ap_a, \gamma^a dx_a]=0.\end{equation}
But from Lemma (3) if $[\tilde{\partial_s}W,\tilde{ds}]=0$, there exists a $\psi=\psi(W)$ such that
$$(\tilde{\partial_s}\psi)\xi(p)=(\partial_s\psi)\xi(p).$$
The result follows.
\\

\noindent \textbf{Remark:} We refer to Equation (12) as a
generalized Dirac equation associated with a curve, and Equation(13) as a generalized Dirac equation associated with geodesics.
It reduces to the usual form of the Dirac equation if we let $\psi=Ae^{\kappa W}$, where $A$ is an arbitrary constant and
$\kappa =\frac{i}{\hbar}$:
\begin{cor} Let $\psi=Ae^{\kappa W}$ where $W=-ms+k$ as defined in
Cor. 1 and $\kappa =\frac{i}{\hbar}$ then
\begin{equation}
\gamma^a\frac{\partial \Psi}{\partial
x^a}=-\frac{i}{\hbar}m\Psi.\end{equation}
\end{cor}
Proof: Direct substitution for $W$ in Equation (13) reduces to the conventional Dirac equation
$$\gamma^a\frac{\partial \Psi}{\partial
x^a}=-\frac{i}{\hbar}m\Psi.$$

\noindent The above equation can be rewritten in the conventional
form, if we multiply across by $-i\hbar \gamma^0$, define
$\alpha^0=\gamma^0$, $\alpha^a=\gamma^0\gamma^{a}$ and let $E=i\hbar H^*=i\hbar\frac{\partial \psi}{\partial t}$ to get
\begin{equation}
\left[-i\hbar\left(\alpha_1\frac{\partial }{\partial x_1} +
\alpha_2\frac{\partial }{\partial x_2} + \alpha_3\frac{\partial
}{\partial x_3}\right) + \alpha_0m\right]\Psi =E \Psi\qquad
\textrm{where} \qquad \Psi=\psi \xi.
\label{dirac}
\end{equation}

\begin{cor} Consider motion in a plane with Hamilton-Jacobi equation $W=-m_os+c_1x^1+c_2x^2+c_3x^3+c_0x^0+d=W_{\|}+W_{\bot}$.
 Let $S_a\equiv \frac{\partial W_{\bot}}{\partial x^a}=c_a$ (called the spin), $\tilde{d\sigma}=\gamma_adx^a+\gamma_a(c^a/m_o)ds$,
$\tilde{\partial_{\sigma}}W\equiv \gamma^a\frac{\partial W}{\partial x^a}=\gamma^a(p_a+c_a)$, then there exists an eigenfunction $\xi$ such that
\begin{equation}(\tilde{\partial_{\sigma}}\psi)\xi(p)=(\partial_{\sigma}\psi)\xi(p).\end{equation}
\end{cor}
Proof: By construction $[\gamma^a\frac{\partial W}{\partial x^a}, \tilde{d\sigma }]=0$ therefore by Lemma 3
$$(\tilde{\partial_{\sigma}}\psi)\xi(p)=(\partial_{\sigma}\psi)\xi(p).$$

\noindent {\bf Remark:} (1) This corollary confirms Martin Rivas' observation that the Dirac equation presupposes the motion of a particle ``in a plane
orthogonal to $S$, which is constant in this frame,'' \cite{rivas} where $S$ is the spin.\\

\noindent (2) The duality relating line increments and the Dirac equation brings to the foreground philosophical issues regarding the difference between
quantum and classical mechanics, and in particular how the two might be related. A more detailed discussion of this point will be found in the next section and also in \cite{ohara}.\\

Next, we prove a theorem that relates the Lie derivative and the
Dirac equation for a particle in a closed system.
\begin{thm} Consider a family of curves $\{\sigma(s)\}$ parameterized
by  the curve length $s$ with unit tangent vectors
$u^a=\frac{dx^a}{ds}$ associated with the increments $ds^2=dx^adx_a$ then
the Lie derivative ${\mathcal{L}}_u({\bf p})=0$ iff there exists a Hamilton-Jacobi function $\psi(W)$ such that $[\gamma^a\frac{\partial
(W)}{\partial x^a},\tilde{ds}]=0$ and 
$$(\tilde{\partial_s}\psi)\xi(p_{\|})=(\partial_s\psi)\xi(p_{\|}).$$
\end{thm}
Proof:($\Rightarrow$) Since ${\mathcal{L}}_u({\bf p})=0$ there exists a
coordinate system (in fact the rest frame) \cite{pois} such that $u^a=\delta^a_0$ and
$p^a_{,0}=0$. Therefore, in a general Lorentz frame $p^a=m_ou^a + m_ov^a$ where $m_o$ is constant and
$\frac{\partial v^a}{\partial x^b}\frac{\partial x^b}{\partial s}=0$. Denote $m_ou^a$ by $p_{\|}^a$.
Then $p^a_{\|}=\frac{\partial W}{\partial x^a}$ defines a Hamilton-Jacobi
function by Cor. 2 and consequently for $\psi=\psi(W)$,
$[\tilde{\partial_s}\psi,\tilde{ds}]=0$. It follows from Lemma 3 that
$$(\tilde{\partial_s}\psi)\xi(p_{\|})=(\partial_s\psi)\xi(p_{\|}).$$

\noindent ($\Leftarrow$)  Recall $\psi=\psi(W)$. Therefore
$\frac{\partial \psi}{\partial x^a}=\psi^{\prime}\frac{\partial
W}{\partial x^a}.$ Now $[\gamma^a\frac{\partial W}{\partial
x^a},\tilde{ds}]=0$ implies $\gamma^a\frac{\partial W}{\partial
x^a}=m(s)\frac{\tilde{ds}}{ds}$ for some function $m(s)$. But
$(\tilde{\partial_s}\psi)\xi(p_{\|})=(\partial_s\psi)\xi(p_{\|})$ implies
$m(s)=\frac{\partial W}{\partial s}$, while Cor. (2) implies that $m(s)=m_o$.
Now let $p_{\|}^a=\frac{\partial W}{\partial x^a}=m_o\frac{dx^a}{ds}$.
Define $p^a=p_{\|}^a+m_ov^a$, where $u^a$ is a Killing vector for $v^a$ (i.e.${\mathcal{L}}_u({\bf v})=0$).
It now follows that
\begin{eqnarray*} {\mathcal{L}}_u(p^a)&=&{\mathcal{L}}_u(p_{\|})+{\mathcal{L}}_u(mv^a)\\
&=&(p_{\|}^a)_{;b}u^b-u^a_{;b}p_{\|}^b\\
&=&mu^a_{;b}u^b-mu^a_{;b}u^b\\
&=&0.\end{eqnarray*} The theorem has been proven.
\\

\noindent \textbf{Remark:} The theorem states that the Dirac
equation associated with a particle exists and is defined locally if
and only if the wave function is Lie transported along the curve
whose action is $W$.
\\

By way of concluding this section, we make some final observations:
\begin{itemize}
\item In general for any Hamilton-Jacobi function $\psi(W)$ it is possible to define
$\tilde{\partial_s}\psi=\tilde{\partial_s}\psi_{\|}+\tilde{\partial_s}\psi_{\bot}$ such that
$[\tilde{\partial_s}\psi_{\|},\tilde{d}s]=0$ and $\{\tilde{\partial_s}\psi_{\|},\tilde{d}s\}=0$.
$\tilde{\partial_s}\psi_{\|}$ is the projected cosine along $\tilde{d}s$ and satisfies
$$(\tilde{\partial_s}\psi)_{\|}\xi(p)=(\partial_s\psi)_{\|}\xi(p),\qquad \textrm{with}
\ p_a=\frac{\partial W}{\partial x^a},$$
where $\tilde{d}s\xi(p)=ds\xi(p)$.
Also, $\tilde{\partial_s}\psi_{\bot}$ defines the spin along $\tilde{ds}$.

\item The Hamilton-Jacobi function can be re-written in covariant form
for a general coordinate system as follows:
\begin{equation} dW=g^{\mu \nu}p_{\mu}dx_{\nu},\end{equation} with
the corresponding wave operator
\begin {eqnarray} \tilde{\gamma}^{\mu}\frac{\partial \psi}{\partial
x^{\mu}}\xi &=&\tilde{\gamma}^{\mu}p_{\mu}\psi^{\prime} \xi \end{eqnarray}
associated with the action along a curve, provided
$2g^{\mu\nu}=\tilde{\gamma}^{\mu}\tilde{\gamma}^{\nu}+\tilde{\gamma}^{\nu}\tilde{\gamma}^{\mu}$, where $\tilde{\gamma}^{\mu}=\frac{\partial x^{\mu}}{\partial x^a}\gamma^a$.

\item The generalized Dirac equation
\begin {eqnarray} \tilde{\gamma}^{\mu}\frac{\partial \psi}{\partial
x^{\mu}}\xi&=&\frac{\partial \psi}{\partial s}\xi. \end{eqnarray}
can be defined along an arbitrary curve and is always covariant.

\item Gauge potentials of the form $A^{\mu}$ can be introduced into the system by defining
\begin{equation} p_{\mu}=m_o\frac{dx_{\mu}}{d\tau}=\partial_{\mu}W-eA_{\mu}.
\end{equation}
However, in general $\oint A^{\mu}dx_{\mu}\neq 0$ and therefore is not an exact differential and $\tilde{ds}\tilde{\partial_s W} \neq \frac{d \psi}{ds}ds$ but does obey Eqn (5).
An analysis of this will be given elsewhere.

\item As already noted, the above approach deepens our understanding of the
Principle of Complementarity. The particle properties should be
directly associated with the path increments. The wave properties emerge from
Equation (13).
\item It should be clear that the strict
form of the Dirac equation (13) pertains to the kinematics and not
the dynamics of the motion. It describes the kinematics with respect
to a local tetrad. The dynamics requires further work, which we will
do in the next section. Indeed, the restriction of the motion to
geodesics also explains why we obtain distinct energy and momentum
levels. Geodesic motion presupposes constant momentum and energy, which can either be discrete or form a continuum depending
on the boundary conditons.
Quantum mechanics associates these constants with quantum numbers.
\end{itemize}

\subsection {Relationship between Quantum and Classical Mechanics}

From a strictly mathematical perspective a particle of rest mass $m_o$ moving in Minkowski space
could equally be a cannon ball (a classical object) or an electron being ejected by a neutron in beta decay (a quantum mechanics object).
In effect, the difference between the two objects are determined by the initial boundary conditions.  In the classical case the line increment from which
the Hamilton-Jacobi function is derived gives rise to a dual non-quantum ``wave-function,'' which is a point mass, given by
$\psi(s)=A\delta (ms)$. In contrast, in the case of quantum mechanics the same line increment is dual to a family of $L^2$ functions
in such way that the initial boundary conditions coming from the physics are statistical, non-deterministic in nature and incorporates quantization.
In other words, in the case of a  strictly classical particle, the mechanics can be determined (in principle) from the initial conditions applied
directly to the properties of the line increments, with the non-quantum ``wave-equation'' representing a point-mass and not contributing any additional information.
In the case of a quantum particle the opposite appears to be true. It is precisely the ``wave-equation'' that encapsulates the dynamics of the particle,
although the solution to the ``wave-equation'' is dependent upon the line increments associated with the classical particle. This also gives a new insight
into the Principle of Complementarity.\\

\noindent \textbf{Example:} Consider a particle of rest mass $m_o$ moving with uniform velocity $u_o$ with respect to proper time along the x-axis in
Minkowski space. The Hamilton-Jacobi function is given by $W=m_ou_xx -m_o\dot{t}t=-m_os$ such that $p_x=m_ou_x$ and $H=m_o\dot{t}$. Indeed as a classical
particle with $x=0$ when $s=0$, then $x=u_os$, where $u_o=\frac{dx}{ds}$. In terms of the coordinate system $(x,t)$ of the laboratory frame this can be
written as $x=u_xt$, where $u_x=(u_o\frac{ds}{dt})$ is constant. Moreover, this information can also be encapsulated in a family of Dirac delta functionals defined
by $\psi(W)=\delta_s(W)\equiv W(s)$ which in term of the laboratory frame is equivalent to $\psi(W(x,t)=\delta(x-u_xt)$. In other words the wave function
as a point mass indicates that at any time $t$ we can find the particle in the position $x=u_xt$ with probability 1. The wave functional per se adds no new
information. It essentially encapsulates the identical information already obtained by analyzing the Hamilton-Jacobi equation derived directly from the
line increments $ds^2=u_odx-\dot{t}dt$.

On the other hand, when we turn to the problem of decay and an electron moving along the x-axis, we have no way of knowing where it will be unless
we place a detector somewhere enroute, which then measures position at a particular instant. Unlike the classical problem, such a measurement does not
allow us to predict subsequent motion in a deterministic way. Rather, we assume that its motion is described by the generalized Dirac equation with eigenvector solution $\psi(W)=A\exp(k(u_ox-\dot{t}t))$, $k$ a dimension full constant. This in turn prompts the question of interpretation. The author claims that such interpretations are multiple and depends on the question being asked.

For example, for a detector placed at $x_o$ the arrival times of a beta-decay source can be modeled with an exponential distribution $\theta \exp(-\theta t)$ with mean $\frac{1}{\theta}$. Moreover, for $x=x_o$, $\psi^2_{x_o}(W)=A\exp(2ku_ox_o)\exp(-2k\dot{t}t)$ defines a probability distribution in $t$. This can be identified with the exponential distribution by letting $\theta= 2k\dot{t}=A\exp(2ku_ox_o)$. In practise, $\theta$ can be measured experimentally by considering an ensemble of $n$ beta particles and measuring the mean time of detection of each one of them at $x_o$. Specifically if the ith decay is detecting at time $t_i$, then $$\frac{1}{\tilde {\theta}}=\frac{1}{n}\sum_{i=0}^n (t_i-t_{i-1})=\frac{t_n-t_o}{n}.$$\\

As a second example, consider n-identical and independent particles in Minkowski space with Hamilton-Jacobi function $W=\sum_{i=1}^n m_os_i$,
(each $i$ representing a different particle). Then the function defined by
\begin{eqnarray} \psi(s_1,s_2,\dots, s_n)&=&\prod_{i=1}^n \psi(s_i)\\
&=&A\exp(k\sum_{i=1}^n (t^2-p^2_1-p^2_2-p^2_3)\\
&=&A\exp(knt^2)\exp\left(-k\sum_{i=1}^n (p^2_1+p^2_2+p^2_3)_i\right)\\
&=&A\exp(knt^2)\exp\left(-k\sum_{i=1}^n \bf{p}^2_i\right),
\label{eqjk}
\end{eqnarray}
where ${\bf p}_i=(p^2_1+p^2_2+p^2_3)_i$, is a solution to the generalized Dirac equation (13), and may be loosely referred to as a ``wave-function.''
\noindent As a function in Minkowski Space it cannot be normalized, as seen from Equations (23)-(25). However, for any fixed $t$, it can be normalized as an $L^p$ function, $p>0$ in the Euclidean space $E^3$. Moreover, as defined it can be used to represent either classical or quantum ideal gases according to the initial boundary conditions imposed on it.\\

For example, for a suitable choice of $A$ and $k$, with $\epsilon_i\equiv \frac {{\bf p}^2_i}{2m_o}$, the squared-``wave function'' can be written as $$\psi^2(W) =A(N)\psi^2(t)\exp\left(-\frac {\beta}{2m_o} \sum_i{\bf p}^2_i\right)=A(N)\psi^2(t)\exp\left(-\beta\sum_i \epsilon_i\right)$$ such that
$$\psi^2(W({\bf p}_1\dots {\bf p}_n)\equiv A(N)\exp\left(-\frac {\beta}{2m_o} \sum_i{\bf p}^2_i\right)=A(N)\exp\left(-\beta\sum_i \epsilon_i\right)$$
is independent of $t$ and defines a Maxwell-Boltzmann distribution for an ensemble of $n$ classical particles defined in the center of mass frame.\\

In addition, if we also require that the values of
${\bf p}^2_i$ are such that $$\epsilon_i={\bf p}^2_i \in \{\hbar^2 k^2_l/2m_o|l=0,1, 2, 3, \dots \},\qquad E=\sum n_l\epsilon_l,\ \sum n_l=n,$$ and impose indistinguishability conditions (encapsulated in the probability amplitude term $A$)  then
$$\psi^2(W({\bf p}_1\dots {\bf p}_n)) =A(N,n_1,n_2, \dots ,n_\infty)\exp\left(-\beta \sum n_i\epsilon_i\right)$$
defines a Bose-Einstein statistic or Fermi-Dirac statistic depending on whether the occupation numbers $n_l$ range over the set $\{0,1,2, \dots \}$ or the set $\{0,1\}.$

The above formalism also begs the question as to why $L^2$ functions are needed. Indeed, for the Maxwell-Boltzmann statistics as derived above,
it would be sufficient to work with any $L^p$ function, provided it is also a solution to the Generalized Dirac equation. However, when we refer to
the Dirac equation proper (\ref{dirac}) the appearance of $i$ and the imposition of periodic boundary conditions to obtain standing waves,
requires that the normalization process be restricted to $L^2$ functions.
In reality, quantum mechanics is an empirical science with measurements and observations being made in real time in the laboratory frame,
and any probability interpretation should be made with this in mind. In this regard, the above examples highlight the importance of boundary conditions
when interpreting the significance of a wave function $\psi$. In the case of the beta decay problem, the wave function $\psi(x,t)\equiv \psi(x)\psi(t|x)$
such that $||\psi(t|x)=1||$. In other words, for each $x$ we can associate a (conditional) probability distribution with $\psi(t|x)$.
Similarly, in the case of both the Maxwell-Boltzmann and Bose-Einstein statistics, $||\psi(W({\bf p}_1\dots {\bf p}_n))=\psi(E)=1||$.\\

\section {Non-geodesic Motion and the Hamiltonian}

In the previous section we related the Hamilton-Jacobi
characteristic function directly to the solution of the generalized Dirac equation and noted that it can
represent both classical and quantum solutions of the equation, dependent upon the initial conditions.
Moreover, from a mathematical perspective $\psi$ can be a functional on a space of compact support $C^{\infty}_c$, or an $L^p$ function defined on some domain. However, within the context of General Relativity, Theorem 1 shows that the general form of the solutions
are determined only locally and not globally, especially
when we consider motion along a non-geodesic. Indeed, the existence
of non-geodesics suggests that other factors other than gravity may be involved. For the purpose of quantum mechanics, we will
take $\psi \in L^2(E^3)$, where $E$ is a Euclidean space and $\Psi \in L^2 \otimes H$ where $H$ is a
finite dimensional Hilbert Space associated with the spin. For example, $\psi \in L^2$ but $\psi \otimes \xi \in L^2\otimes H$. $\Psi$ and $\psi$ will also be referred to as defining the state of the system, with and without spin respectively.

In this regard, we need also to be aware that from a strict physics perspective, things are more nuanced. Dirac, for example,
defines the state as the maximum knowledge that we may have about the system. However,
when one formulates a mathematical theory one is always limited by the definitions and restrictions imposed by the ``space'' within which the
theory is formulated.  For example, in the previous section $W$ was restricted to being a Hamilton-Jacobi function
which required that momentum be defined as $p_a=\frac{\partial W}{\partial x^a}$. However, from Lemma 1, we know that $\psi(kW)$ is also a Hamilton-Jacobi function and in particular $\frac{\partial \psi(kW)}{\partial x^a}=kp_a\psi(kW)$, if $\psi$ is an eigenfunction. Moreover, if $k$ is then chosen to be the complex number $i$, quantum bound states result. Similarly, the Hamiltonian $H$ was defined strictly in terms of position
and momentum. Consequently, in this context the energy states of the system depend only on position and momentum and on the imposed boundary conditions.
In other branches of physics, the state, may also depend on temperature, potential, electric and magnetic fields.

\subsection{Hamilton's Equations of Motion}

The Hamilton-Jacobi equation is given by $W=\int ({\mathbf pdx}-Hdt)$ with the understanding that the integration is independent of the path. Using this, we now derive Hamilton's equations of motion directly from this definition, without making any explicit recourse to the Calculus of Variations.

First, for each differentiable function $W$, we write $\frac{dW}{ds}=\dot{W}$. Also note that $||\frac{dx^a}{ds}||=1$ requires that $\dot{W}=\frac{\partial W}{\partial s}=-H(s)$, and on taking the derivative of the Hamilton-Jacobi function with respect to s we obtain
\begin{equation} -H(s)={\mathbf{p\dot x}}-H\dot t,
\end{equation}
with the understanding that for $a\in \{1,2,3\}$
\begin{eqnarray}\dot{W}=\frac{\partial W}{\partial s}=-H(s),\qquad \frac{\partial W}{\partial x^a}=p_a\qquad
\textrm{and}\qquad \frac{\partial W}{\partial t}=-H.
\end{eqnarray}
Note that these can be written in covariant form $p^{\mu}=g^{\mu \nu}\frac{\partial W}{\partial x^{\nu}}$. However, for simplicity and clarity we will continue to work with local tetrads.

Differentiating (27) with respect to $s$ gives
\begin{equation}\frac{\partial H}{\partial s}=\dot{H}(s) \qquad \textrm{and}\qquad \dot{p}^a=-\frac{\partial H(s)}{\partial x^a}.
\end{equation}
To derive the remaining equation of motion we follow a method introduced by Synge and Griffith(see \cite{Synge}):
\begin{eqnarray}
\dot{p}_a&=&\frac{\partial}{\partial x^a}\left(\frac{dW}{ds}\right)\\
&=&\frac{\partial}{\partial x^a}\left(\frac{\partial W}{\partial x^b}\dot{x^b}+\frac{\partial W}{\partial t}\dot{t}\right)\\
&=&\frac{\partial^2 W}{\partial x^a \partial x^b}\dot{x^b}+\frac{\partial^2 W}{\partial x^a \partial t}\dot{t}.
\end{eqnarray}
Also,
\begin{eqnarray}
\frac{\partial}{\partial x^a}\left(\frac{\partial W}{\partial t}\right)\dot{t}
&=&-\frac{\partial }{\partial x^a}\left(H\left(x^b,\frac{\partial W}{\partial x^b},t\right)\right)\dot{t}\\
&=&-\frac{\partial H}{\partial x^a}\dot{t}-\frac{\partial H}{\partial p^b}\frac{\partial^2 W}{\partial x^a \partial x^b}\dot{t}\\
&=&-\frac{\partial H(s)}{\partial x^a}-\frac{\partial H}{\partial p^b}\frac{\partial^2 W}{\partial x^a \partial x^b}\dot{t}.
\end{eqnarray}
Combining equations (31) and (34) and substituting from (28) yields
\begin{equation} \frac{\partial^2 W}{\partial x^a \partial x^b}\left(\dot{x}^b-\frac{\partial H(s)}{\partial p^b}\right)=0.\end{equation}
Therefore,
\begin{equation} \dot{x}^b =\frac{\partial H(s)}{\partial p^b}\qquad \textrm{provided}\ det\left(\frac{\partial W}{\partial x^a \partial x^b}\right)\neq 0.\end{equation}
This completes the derivation of the canonical equations of motion.

In terms of tetrad summation notation these can be rewritten as
\begin{equation}
\frac{dx^a}{ds}=\eta^{ab}\frac{\partial H(s)}{\partial p^b}, \qquad
\qquad \frac{dp^a}{ds}=-\eta^{ab}\frac{\partial H(s)}{\partial x^b}.
\end{equation}
These are the
same equations assumed by Horwitz, Schieve and Piron in their work on Stueckelberg
theory applied to the Gibb's ensemble\cite{hor}. It should also be noted, in reference to Eqn. (37) that in tetrad coordinates by the principle of equivalence $\frac{dp^a}{ds}=\frac{Dp^a}{ds}$, where $D$ represents the covariant derivative. This follows because the affine connection vanishes on a geodesic. In terms of a generalized coordinate
system both equations of motion can be subsumed into the covariant form:
\begin{equation}
\frac{dx^{\mu}}{d\tau}=g^{\mu\nu}\frac{\partial K}{\partial
p^{\nu}}, \qquad \qquad
\frac{Dp^\mu}{d\tau}=-g^{\mu\nu}\frac{\partial K}{\partial x^{\nu}},
\end{equation}
where $\frac{Dp^{\mu}}{d\tau}=p^{\mu}_{;}u^{\nu}=\dot{p}+\Gamma^{\mu}_{\nu \lambda}\dot{x}^{\mu}\dot{x}^{\lambda}$. Note that in general the covariant derivative is necessary on the left hand side, since $\frac{dp^{\mu}}{ds}$ is not a tensor and transforms like acceleration under a change of coordinate systems. On the other hand the right hand side of the second equation is already covariant.

\subsection {Non-commutative Operators}
We begin this section by recalling that for this article $\Psi=\psi^ie_i$ represents a spinor on a manifold, while
$\psi^i$ are scalar functions (cf. page 2). Also if $\Psi=\psi\xi$, where $\xi$ is a spinor
then $\psi$ is said to define a scalar field. It then follows that
$\tilde{\partial}_s\psi(W)\equiv \gamma^a\frac{\partial \psi(W)}{\partial x^a}$ and $\tilde{\partial}_s\psi(H)=\gamma^a\frac{\partial\psi(H)}{\partial x^a}$
are operators, because of the presence of the $\gamma$ matrices, where $W$ is the Hamilton-Jacobi function and $H$ is the Hamiltonian of the system.
In particular,
\begin{equation} [\tilde{\partial}_s\psi(W), \tilde{\partial}_s\psi(H)]\ne
0.\end{equation}
To see this, note that from equations (27) and (28) applied to $\psi(W)$ and $\psi(H)$ respectively,
we obtain
\begin{equation}\tilde{\partial}\psi(W)=\gamma^a\frac{\partial
\psi(W)}{\partial x^a}= \gamma^ap_a\psi^{\prime}(W)\end{equation}
and
\begin{equation}\tilde{\partial}\psi(H)=\gamma^a\frac{\partial
\psi(H)}{\partial x^a}=
\gamma^a\dot{p}_a\psi^{\prime}(H).\end{equation} Now unless
$\dot{p}^a=g(t)p^a$ for some function $g(t)$ then they cannot
commute. Indeed, the condition $\dot{p}^a=g(t)p^a$ is equivalent to
$x^a=\exp( \int^t g(\omega)d\omega)$ independently of $a$, which
defines a non affine parameter along a geodesic (cf. \cite{pois} Prob., 1.13, n2).
Consequently, there
do not exist simultaneous eigenvectors, and both operators cannot be
simultaneously measured, except along a geodesic. This in itself
explains some of the difficulty with dynamics in quantum mechanics. It is also clear from Equation (5) that the non-commutativity is very much related to
the presence of the $\gamma$ matrices and the Dirac Algebra, and although $\psi(W)$ or $\psi(H)$ may themselves be $L^p$ functions nevertheless $\Psi=\psi\xi$, $\xi$ a spinor, may be interpreted as an element of the finite four dimensional Hilbert space associated with the spin of the particle. Indeed, the non-commutativity indicates that spin only becomes manifest in accelerations, and also suggests a second
quantization procedure associated with particles moving off geodesics.

In practice, both equations can be seen as useful depending on the circumstances. When the dynamical system corresponds to motion on geodesics then $\dot{p}=0$ and has no role to play. In this case the generalized Dirac equation (13) can be used. In the event that the dynamical system is undergoing accelerations with respect to the laboratory frame then we could solve the eigenvector equation
\begin{equation}\tilde{\partial}\psi(H)\xi=\gamma^a\frac{\partial
\psi(H)}{\partial x^a}\xi=\frac{dm}{ds}\xi
\end{equation}
where $\frac{dm}{ds}=||f^a||$ is the norm of the four-force $f^a$. It should also be pointed out that since the Principle of Equivalence guarantees the existence of a coordinate system $x^a$ such that $\ddot{x^a}=0$, the existence of absolute acceleration presupposes the existence of a non-gravitational force such that $a^a=\ddot{x}\neq 0$. This also means that in the presence of such an acceleration the rest mass will not be a constant but will change according to whether its accelerating or decelerating. Accordingly, depending on whether $\frac{dm}{ds}=0, >0, <0$, three cases arise which could be interpreted as defining a null, time-like, and space-like event in momentum space. This, together, with the second quantization properties will be investigated in another research paper.

\subsection{Statistical Mechanics and Ideal Gases}

Another interesting application is to theory of Ideal gases.  Indeed, as noted in Section 2.1 both classical and quantum statistics distributions can be derived from solving the generalized Dirac equation, with the difference between the two being related to the boundary conditions. In fact, in equations (23)-(25) we already assumed the form of the ``wave-function'' for an ideal gas, both classical and quantum. However, it now remains to derive this fundamental Hamiltonian function as an application of Hamilton's equation's of motions.

First, consider a single particle with a Hamilton-Jacobi function given by $W=-ms$ associated with motion along a geodesic in Minkowski space, then $H=\frac{\partial W}{\partial s}=m$. In terms of the equations of motion there exists a differentiable function $\psi=\psi(H)$ such that
\begin{equation} \frac{\partial \psi}{\partial
x^a}=-\psi^{\prime}\dot{p}^a,\quad {\rm where}\ \dot{p}=\frac{dp}{ds}, \end{equation}which can be re-written
in spinor notation as
\begin{equation}\gamma^a \frac{\partial \psi}{\partial
x^a}=-\gamma^a\psi^{\prime}\dot{p}^a. \end{equation} Taking the
dot product of (43) with $m\frac{dx^a}{ds}$ and using the chain rule,
(or equivalently by taking the inner product $\frac{1}{2}\{\gamma^a \frac{\partial
\psi(H)}{\partial x^a},\gamma^a p_a\}$ in (44)) gives
\begin{equation}
\frac{d\psi(H)}{ds}=-\psi^{\prime}(H)\dot{p}^ap_a.\end{equation}
To solve, recall that $\psi^{\prime}(H)=\frac{d\psi(H)}{dH}$ and $2\dot{p}_ap^a=\frac{d}{ds}(p_ap^a)$, and consequently $H=\frac12p_ap^a$.
In
particular when $\psi^{\prime}=k\psi$, an eigenvector, and $k$ an eigenvalue, solving
for $\psi$ gives
\begin{eqnarray} \psi = Ae^{\frac{k}{2}p^ap_a}.\end{eqnarray}
If $k$ is real and time dependent then
$$\psi(t, x)=A\psi(t)\psi(x|t)=A\exp(kt^2)\exp(-k(p^2_1+p^2_2+p^2_3)),$$ is not normalizable. However, for each $t$, $\psi(x|t)$ is an $L^p(E^3)$ function and can be normalized. Indeed, the normalized set of functions $\{\psi(x|t)\}$ defines a Markov process in $t$.  Moreover, if $k$ is a constant as it is for geodesic motion, then $\psi(x_1,x_2,x_3|t)=\psi(x_1,x_2,x_3)$. On the other hand, if we take $k=\frac{i}{\hbar}$ then $\psi(p|t)=\psi(p)$ is an $L^2$ function with periodic boundary conditions.

With this in mind, we now extend this result to a system of $n$ independent particles as would occur in some
statistical systems, and define the joint wave function as an independent product of n single particle eigenfunctions with real $k$, or equivalently as an
ensemble of $n$ independent eigenfunctions. Written as an $L^2$ function (although for a classical particle we could equally work with $L^1$)
\begin{eqnarray} \psi = e^{\frac{k}{2}\sum^n_1p^ap_a}.\end{eqnarray}
Moreover, if the motion is along a geodesic then $H=constant$ and
$p^ap_a=m$ for each particle Consequently,
\begin{eqnarray} \psi=
c\exp\left[\frac{k}{2}m\sum_n(\dot{t}^2-\dot{x}^2_1-\dot{x}^2_2-\dot{x}^2_3)\right]=e^{\frac{k}{2}nm}.
\end{eqnarray} Defining $T=\frac{1}{k_Bk}$, to be the temperature, where $k_B$ is Boltzmann's constant, and using the separation of variables for time independent states to write
$\psi=\psi(t)\psi(x_1,x_2,x_3)$, where
\begin{equation}
\psi(x_1,x_2,x_3|t)=c\exp\left({\frac{-\frac{1}{2}m}{k_BT}\Sigma_n(\dot{x}^2_1+\dot{x}^2_2+\dot{x}^2_3)}\right)\end{equation}
determines a Maxwell-Boltzmann statistics for free particles at any time $t$.
Note also that $|\psi(x_1,x_2,x_3|t)|^2$ defines a normal distribution provided the variance is given by $\sigma_x^2=\sigma_y^2=\sigma_z^2=\sigma^2=\frac{k_BT}{2m}$.  Equations (47) and (48) can be interpreted to mean that the system is in
equilibrium, with total conserved energy $\sum_n m$. Indeed, for each (local) $t$
the same Maxwell distribution occurs. Moreover,
if T is not constant then $k=k(T)$ varies and in this case, on
solving for $\psi$ in (43) one obtains
\begin{eqnarray} \psi = e^{\int k(T)p_a\dot{p}^a dt}.\end{eqnarray}
Also from equation (44) we should note that
$\frac{\psi}{\psi^{\prime}}$ is always an exact differential, and
incorporates $k(T)$. If $k=k(T)$ has no explicit time dependence
then $\psi$ will define a stationary state.  Note that in the case of a closed system, $T$ is proportional to the population variance of the velocities
$\sum_n\frac{(u_i-\mu)^2}{n}$.

\section{Conclusion} The article has attempted to establish a
relationship between the metrics of General Relativity and Quantum
Mechanics. This has been achieved by first relating the metric
structure of spacetime to the Hamilton-Jacobi function and then
using this relationship to derive a Generalized Dirac equation. In addition we have derived Hamilton's equations of motion directly
from the Hamilton-Jacobi equation and then used these equations to determine a dynamical
equation for the evolution of the system.  The dynamical
equations derived do not commute with the generalized Dirac equation
and consequently cannot be measured simultaneously. The dynamical
equations also permit the derivation of the statistical mechanics of
the system. Indeed, in this paper the Maxwell-Boltzmann distribution
was derived directly from the equations of motion.

Finally, it should be noted that we have restricted ourselves to
scalar fields as defined in the introduction. However, in its most
general form, we can write
\begin{equation}\gamma^a\frac{\partial
\Psi}{\partial x^a}=\Phi\end{equation} where $\Phi$ would be defined
by the physics of the problem. For example, Maxwell's equations in
Minkowski space can be written in spinor form as
\begin{equation}i\alpha^a\frac{\partial
\Psi}{\partial x^a}=-4\pi\Phi,\end{equation} where $\phi_0=\rho$ is
charge density, and $\phi_a=j_a,\ a\in\{1,2,3\}$ is a current
density. Also in this case, $\phi_0=0$ and $\phi_a=H_a-iE_a$, where
$H_a$ and $E_a$ are the magnetic and electric fields respectively
\cite{moses}. Such cases would need further study.
\\

\noindent \textbf{Acknowledgement:} I would like to thank Prof.
Lamberto Rondoni from the \emph{Politecnico di Torino }and also the referees for their
invaluable suggestions and input while I was writing this paper.

%
%

%

\begin{thebibliography}{99.}
%
%
%
\bibitem {schr}O'Raifeartaigh L (1997)
The Dawning of Gauge Theory. Princeton University Press, Princeton
New Jersey: 92.
\bibitem {pois}Poisson, E (2004) A Relativistist's Toolkit.
Cambridge University Press, Cambridge: 9.
\bibitem{Synge} Synge J\& Griffith G (1959) Principles of
Mechanics. Mc-Graw Hill New York and Tokyo: 448-450.


\bibitem{marie} Celerier, Marie-Noell \& L. Nattale (2003)
Electromagn. Phenom: 3-70-80.
\bibitem{ng} Ng, Y. and H. van Dam (2003) Phys.Lett A309.
\bibitem{hor} Horwitz L. and W. Schieve (1981) Ann. of Phys.
137,307.
\bibitem{moses} Moses H (1958) Physical Rev. 6:1670-1679.
\bibitem{ohara} O'Hara P (2005) Found. Phys. 35:1563--1584.
\bibitem{rivas} Rivas, M. (1994) J. Math. Phys., Vol 35, No. 7.
\bibitem{tip} Tipler, F. (2010) arXiv: 1007.4566v1[quant-ph].

\end{thebibliography}
%



\end{document}